\documentclass{article}
\usepackage{amsmath}
\usepackage{graphicx}
\usepackage{hyperref}
\usepackage{geometry}
\usepackage{graphicx}
\usepackage{geometry}
\usepackage{array}
\usepackage{tikz}
\usetikzlibrary{shapes.geometric, arrows, positioning}
\usepackage{natbib} 

\title{PPINtonus: Early Detection of Parkinson’s Disease Using Deep-Learning Tonal Analysis}
\author{Varun Reddy\\ Academy of Engineering and Technology}
\date{6/2/2022}

\begin{document}

\maketitle

\begin{abstract}
PPINtonus is a system for the early detection of Parkinson's Disease (PD) utilizing deep-learning tonal analysis, providing a cost-effective and accessible alternative to traditional neurological examinations. Partnering with the Parkinson's Voice Project (PVP), PPINtonus employs a semi-supervised conditional generative adversarial network to generate synthetic data points, enhancing the training dataset for a multi-layered deep neural network. Combined with PRAAT phonetics software, this network accurately assesses biomedical voice measurement values from a simple 120-second vocal test performed with a standard microphone in typical household noise conditions. The model's performance was validated using a confusion matrix, achieving an impressive 92.5 \% accuracy with a low false negative rate. PPINtonus demonstrated a precision of 92.7 \%, making it a reliable tool for early PD detection. The non-intrusive and efficient methodology of PPINtonus can significantly benefit developing countries by enabling early diagnosis and improving the quality of life for millions of PD patients through timely intervention and management.
\end{abstract}

\section{Background}

    Parkinson's Disease (PD) is a progressive neurodegenerative disorder that primarily affects motor function due to the loss of dopamine-producing neurons in the substantia nigra, a region of the brain. The cardinal motor symptoms of PD include tremors, rigidity, bradykinesia (slowness of movement), and postural instability. These symptoms significantly impact the quality of life of individuals and typically worsen over time. In addition to motor symptoms, non-motor symptoms such as cognitive impairment, mood disorders, and sleep disturbances also occur, further complicating disease management. \\

    \noindent Early detection of PD is crucial as it allows for the timely initiation of therapies that can alleviate symptoms and potentially slow the disease progression. However, early-stage PD is often challenging to diagnose because the symptoms can be subtle and may overlap with other conditions. Traditional diagnostic methods for PD involve clinical evaluations, including neurological examinations and the patient's medical history. While these methods are effective, they rely heavily on the expertise of medical professionals and are subjective to some extent. \\
    
    \noindent Advanced imaging techniques such as computed tomography (CT) and magnetic resonance imaging (MRI) are also used to aid in the diagnosis of PD. These methods provide detailed images of the brain, allowing for the identification of structural abnormalities \cite{theodoros2008}. However, the primary limitation of these techniques is their high cost and the need for specialized equipment and trained personnel \cite{ogbole2018}. This makes them less accessible, especially in developing countries with limited healthcare resources. To address the need for more accessible diagnostic tools, researchers have explored various biomarkers that can be indicative of PD. Among these, vocal biomarkers have gained significant attention. PD affects the muscles involved in speech production, leading to changes in voice quality, pitch, loudness, and articulation \cite{sapir2011}. These changes occur early in the disease process, making vocal analysis a promising avenue for early PD detection. ML models have been increasingly applied to analyze vocal biomarkers for PD detection. These models can process large amounts of data and identify patterns that may not be apparent to human observers. Traditional ML approaches have been employed to classify vocal features extracted from speech recordings. Despite the potential of ML models in PD detection, several challenges remain. One major challenge is the variability in vocal features among different individuals, which can be influenced by factors such as age, gender, and the presence of other medical conditions. This variability can make it difficult for models to generalize across different populations \cite{arora2015}. Additionally, traditional ML models often require extensive feature engineering, which involves manually selecting and transforming raw data into a format suitable for model training. This process can be time-consuming and may not capture the full complexity of the data. Deep learning models, particularly neural networks, offer an alternative approach by automatically learning hierarchical representations of data. CNNs and RNNs have been used to analyze speech signals and detect PD \cite{liao2020}. These models can learn directly from raw audio data, reducing the need for manual feature engineering \cite{marti2019}. However, the performance of deep learning models is heavily dependent on the availability of large, annotated datasets. In the context of PD detection, obtaining a sufficiently large and diverse dataset of vocal recordings is challenging, which limits the effectiveness of these models. \\

    \noindent Another approach that has been explored is the use of hybrid models that combine traditional ML techniques with deep learning \cite{miao2019}. These models aim to leverage the strengths of both approaches to improve diagnostic accuracy. For example, a hybrid model might use deep learning to extract features from raw audio data and then apply a traditional ML classifier to make predictions \cite{he2016}. While promising, these models still face data availability and variability challenges.

\section{Methodology}
    \subsection{Data Collection}
        Our study utilized the UC Irvine Parkinson's Disease Detection Dataset as the primary source of biomedical voice measurements. Additionally, we collaborated closely with vocal specialists at the Parkinson's Voice Project and biomedical engineering experts at the Monroe Advanced Technical Academy. The dataset underwent an extensive preprocessing phase to prepare it for effective model training. Initially, the data was thoroughly cleaned to remove any inconsistencies or anomalies that could negatively impact model performance. This cleaning process was followed by one-hot encoding, a technique used to convert categorical variables into a numerical format suitable for machine learning algorithms. \\
        
        \noindent To further enhance our dataset, we generated additional synthetic data using a Conditional Generative Adversarial Network (cGAN) \cite{goodfellow2014}. A cGAN consists of two primary components: a generator and a discriminator. The generator's role is to produce synthetic data samples that resemble real data, while the discriminator evaluates these samples to determine their authenticity. The generator creates synthetic samples conditioned on actual data features, which are then assessed by the discriminator. Both components are trained simultaneously in a competitive setting, where the generator continuously improves its ability to produce realistic data, and the discriminator enhances its capability to distinguish between real and synthetic samples. This iterative process continues until the discriminator can no longer reliably differentiate between real and synthetic data, indicating that the generator has successfully learned to produce high-quality synthetic samples \cite{pang2021}. The synthetic data generated by the cGAN was rigorously validated and subsequently used to augment our training dataset. This approach significantly increased the volume of data available for training, enhancing the robustness and generalizability of our neural network model. \\

        \begin{figure}[h]
            \centering
            \includegraphics[width=0.5\textwidth]{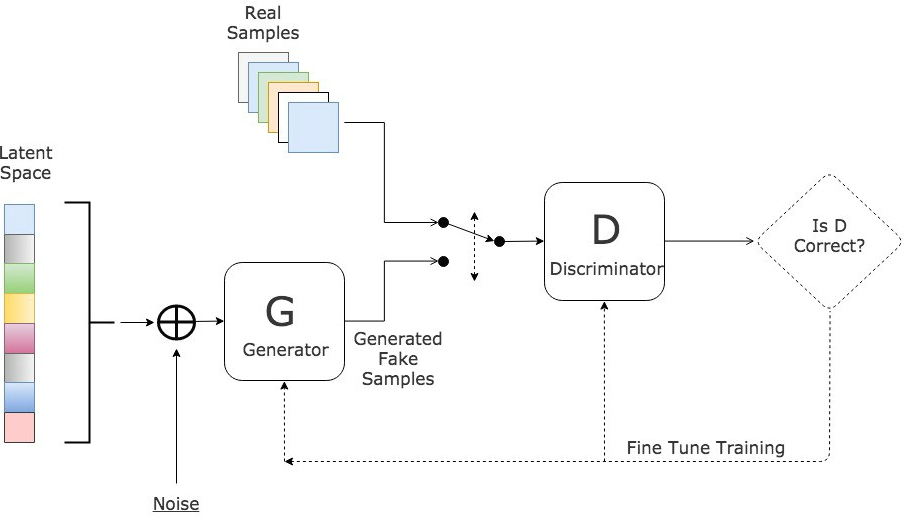}
                \caption{Illustration of a Generative Adversarial Network (Gharakhanian).}
            \label{fig:example}
        \end{figure}

        \noindent By leveraging a diverse set of samples, our model was better equipped to detect Parkinson's Disease across a variety of conditions and patient profiles. In the feature extraction phase, we used PRAAT phonetics software to derive critical vocal features from the dataset. This software provided precise measurements of various vocal characteristics, which were further validated by vocal specialists to ensure their accuracy and relevance in the context of Parkinson's Disease detection.

   \subsection{Deep Learning Methodology}
The neural network architecture designed for this study includes multiple fully connected (dense) layers interspersed with ReLU activation functions. The output layer uses a sigmoid activation function to produce a probability score indicating the likelihood of PD. \\

\noindent One significant challenge in deploying a deep learning model trained on controlled data to real-world environments is handling noise. Training data is typically collected in sound booths using high-quality condenser microphones, ensuring minimal background noise and high fidelity. However, real-world applications, especially in developing countries, often involve standard smartphone microphones in typical household environments, which are prone to various types of noise. To address this, we incorporate several noise-handling techniques. Firstly, we apply data augmentation techniques to simulate real-world conditions during training. This includes adding different types of noise (e.g., white noise, background chatter, household sounds) to the clean audio recordings. By training the model on augmented data, it learns to differentiate between noise and the relevant vocal features indicative of PD. Additionally, we integrate noise reduction algorithms as a preprocessing step before feeding the audio data into the model. We employ deep learning-based denoising autoencoders to reduce background noise, enhancing the clarity of the extracted vocal features \cite{benba2021}. Furthermore, we employ microphone calibration techniques to account for the differences in audio quality between high-end condenser microphones and standard smartphone microphones. We reduce the discrepancy between training and real-world data by calibrating the recordings to match the audio profile of high-quality data. This ensures that the model remains robust and reliable when used in practical settings with standard equipment \cite{dehak2010}. \\

\noindent Raw audio data is initially preprocessed to extract relevant features using PRAAT software. Features such as fundamental frequency (F0), jitter, shimmer, and harmonics-to-noise ratio (HNR) are derived from the audio recordings. Clean training data is then augmented with various types of noise to simulate real-world conditions, and noise reduction algorithms are applied to the noisy data to improve signal quality. The neural network is trained using the augmented and noise-reduced data, with dropout layers applied to prevent overfitting. Finally, performance metrics such as accuracy, precision, recall, and F1-score are calculated and thoroughly analyzed to assess the model's effectiveness. Based on these results, a Bayesian optimization technique is employed to further fine-tune the model's hyperparameters \cite{tsanas2012} \cite{kingma2013}. This method allows for a systematic search of the hyperparameter space, ensuring that the model achieves optimal performance and accuracy by evaluating the trade-offs between different configurations and converging on the most effective parameter set.

\subsection{Real Time PD Detection}
Real-time detection of Parkinson's Disease (PD) through vocal analysis involves selecting the most effective vocal tests to extract reliable Biomedical Voice Measurements (BVMs). These measurements are crucial for the machine learning model to accurately predict the presence of PD. Research indicates that certain types of vocal tasks are more effective at revealing the subtle vocal characteristics affected by PD \cite{hinton2012}. The primary vocal tests used for PD detection include sustained vowel phonations, sentence readings, and complex speech tasks \cite{theodoros2008}. Each test type provides unique insights into different aspects of vocal function affected by PD. Sustained vowel phonations involve prolonged pronouncing vowels such as /a/, /i/, and /u/. This test is simple to administer and has been shown to effectively reveal abnormalities in vocal fold vibration and control. Research indicates that patients with PD exhibit increased jitter (frequency variation) and shimmer (amplitude variation) during sustained phonation due to the reduced ability to maintain a stable pitch and volume \cite{skodda2011}. These measurements are critical BVMs for the model. Sentence readings involve having the patient read predefined sentences aloud. This task assesses more complex speech functions, including prosody (intonation), articulation, and rhythm. Sentences are designed to include a variety of phonemes and stress patterns to challenge the patient’s speech control. Studies have shown that PD patients often have a reduced range of pitch and volume modulation, as well as increased pauses and hesitations \cite{sapir2011}. These features can be quantified as BVMs and used to train the machine learning model. Complex speech tasks include spontaneous speech, narrative tasks, and rapid repetition of syllables (e.g., "pa-ta-ka"). These tasks are more demanding and can highlight the motor planning deficits characteristic of PD. For instance, diadochokinetic rate (the ability to make rapid, alternating movements) can be measured during syllable repetition tasks. PD patients typically show a slower rate and irregular rhythm, which are valuable BVMs for the model. Extensive research has demonstrated the efficacy of these vocal tests in identifying PD-specific vocal impairments. Sustained vowel phonations are particularly useful for their simplicity and sensitivity to vocal tremors and stability issues. Sentence readings provide a broader assessment of speech control and are effective in capturing prosodic abnormalities. While more challenging to administer, complex speech tasks offer comprehensive insights into the neuromuscular control of speech \cite{rusz2011}.

\begin{table}[h!]
    \centering
    \caption{Various Biomedical Vocal Features and Their Descriptions}
    \begin{tabular}{|m{4cm}|m{8cm}|}
        \hline
        \textbf{Feature} & \textbf{Description} \\ \hline
        Fundamental Frequency (F0) & Average pitch of the voice, indicating vocal fold vibration rate. \\ \hline
        Jitter & Frequency variation between cycles, indicating vocal stability. \\ \hline
        Shimmer & Amplitude variation between cycles, indicating vocal amplitude regularity. \\ \hline
        Harmonics-to-Noise Ratio (HNR) & Ratio of harmonic to noise components, indicating voice quality. \\ \hline
        Formant Frequencies (F1, F2, F3) & Resonant frequencies of the vocal tract, crucial for vowel sounds. \\ \hline
        Intensity & Loudness of the voice, reflecting vocal energy. \\ \hline
        Voice Onset Time (VOT) & Interval between consonant release and vocal fold vibration. \\ \hline
        Speech Rate & Speed of speech, indicating motor control of speech production. \\ \hline
        Diadochokinetic Rate (DDK) & Rate of rapid, alternating movements, assessing neuromuscular coordination. \\ \hline
        Pitch Range & Range between the highest and lowest pitches, indicating vocal flexibility. \\ \hline
        Speaking Fundamental Frequency (SFF) & Mean pitch during continuous speech. \\ \hline
        Maximum Phonation Time (MPT) & Longest time a vowel can be sustained, indicating respiratory control. \\ \hline
        Cepstral Peak Prominence (CPP) & Measure of voice quality, higher values indicate clearer voice. \\ \hline
        Voice Range Profile (VRP) & Range of pitch and intensity, indicating vocal capacity. \\ \hline
        Phonation Threshold Pressure (PTP) & Minimum subglottal pressure needed to initiate phonation. \\ \hline
        Amplitude Perturbation Quotient (APQ) & Measure of short-term amplitude variations. \\ \hline
        Normalized Noise Energy (NNE) & Ratio of noise energy to total energy in the voice signal. \\ \hline
    \end{tabular}
    \label{tab:bvm_features}
\end{table}

\noindent A study by Rusz et al. (2011) quantitatively analyzed the speech of early untreated PD patients and found that these patients exhibited significant deviations in fundamental frequency (F0), jitter, shimmer, and harmonics-to-noise ratio (HNR) compared to healthy controls. These deviations were most pronounced during sustained vowel phonations but were also evident during sentence readings and complex speech tasks. Another study by Skodda et al. (2011) highlighted the importance of prosodic features, such as intonation and speech rate, which are best captured during sentence readings and spontaneous speech tasks. The combination of different vocal tasks ensures a comprehensive assessment of the patient’s speech abilities, thereby providing a robust set of BVMs for the machine-learning model. 

\begin{figure}[h!]
        \centering
        \includegraphics[width=0.8\textwidth]{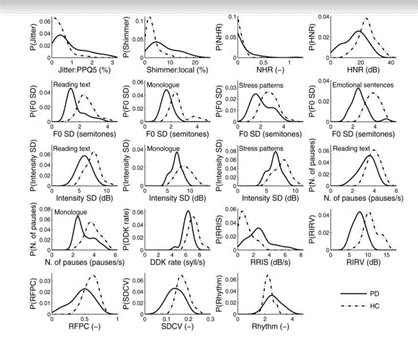}
        \caption{Figure 2: Differences in specific BVMs between Parkinson's patients and healthy individuals (Rusz)}
    \end{figure}

\begin{table}[h!]
    \centering
    \caption{Sample Vocal Test Tasks for Real-Time PD Detection}
    \begin{tabular}{|c|p{4cm}|p{8cm}|}
        \hline
        \textbf{Task Code} & \textbf{Speech Data} & \textbf{Description} \\ \hline
        \textbf{TASK 1} & Sustained phonation of /i/ & At a comfortable pitch and loudness, as constant and long as possible, at least 5 s. \\ \hline
        \textbf{TASK 2} & Rapid syllable repetition & Steady repetition of /pa/-/ta/-/ka/ syllables, repeated at least 5 times on one breath. \\ \hline
        \textbf{TASK 3} & Sustained vowels /a/, /i/, /u/ & Approximately 5-second sustained vowels at a comfortable pitch and loudness. \\ \hline
        \textbf{TASK 4} & Sentence reading & Reading a phonemically balanced text of 136 words. \\ \hline
        \textbf{TASK 5} & Monologue & Speaking for approximately 90 s about a familiar topic (e.g., recent events, interests). \\ \hline
        \textbf{TASK 6} & Stress pattern reading & Reading the same text containing 8 variable sentences of 71 words with varied stress patterns. \\ \hline
        \textbf{TASK 7} & Emotional sentence reading & Reading 10 sentences with specific emotions in a neutral tone, covering various emotional states. \\ \hline
        \textbf{TASK 8} & Rhymed text reading & Reading rhymes of 34 words following the example set by the examiner. \\ \hline
    \end{tabular}
    \label{tab:vocal_tasks}
\end{table}

\newpage

\section{Proposed Model Architecture}

\begin{tikzpicture}[node distance=2.5cm, every node/.style={fill=white, font=\sffamily, align=center},scale=0.85, align=center, transform shape]

\node (start) [rectangle, draw, rounded corners, fill=blue!20, text width=5cm] {Vocal Test Administration through PPINtonus software};
\node (data) [rectangle, draw, rounded corners, fill=green!20, below of=start, text width=5cm] {Data Acquisition via Parselmouth (PRAAT python library)};
\node (preproc) [rectangle, draw, rounded corners, fill=yellow!20, below of=data, yshift=-0.5cm, text width=5cm] {Preprocessing};
\node (nr) [rectangle, draw, rounded corners, fill=orange!20, below of=preproc, xshift=-3.5cm, yshift=-0.5cm, text width=3cm] {Noise Reduction};
\node (da) [rectangle, draw, rounded corners, fill=orange!20, below of=preproc, text width=3cm, yshift=-0.5cm] {Data Augmentation};
\node (mc) [rectangle, draw, rounded corners, fill=orange!20, below of=preproc, xshift=3.5cm, yshift=-0.5cm, text width=3cm] {Microphone Calibration};
\node (feature) [rectangle, draw, rounded corners, fill=red!20, below of=preproc, yshift=-5cm, text width=5cm] {Feature Extraction};
\node (nn) [rectangle, draw, rounded corners, fill=purple!20, below of=feature, yshift=-0.5cm, text width=5cm] {Neural Network Classification Model};
\node (train) [rectangle, draw, rounded corners, fill=cyan!20, below of=nn, yshift=-0.5cm, text width=5cm] {Training Process};
\node (eval) [rectangle, draw, rounded corners, fill=pink!20, below of=train, yshift=-0.5cm, text width=5cm] {Evaluation and Fine-Tuning};

\draw[->] (start) -- (data);
\draw[->] (data) -- (preproc);
\draw[->] (preproc) -- (nr);
\draw[->] (preproc) -- (da);
\draw[->] (preproc) -- (mc);
\draw[->] (nr) -- (feature);
\draw[->] (da) -- (feature);
\draw[->] (mc) -- (feature);
\draw[->] (feature) -- (nn);
\draw[->] (nn) -- (train);
\draw[->] (train) -- (eval);

\end{tikzpicture}

\section{Results}
    The Generative Adversarial Network (GAN) model was trained over 10000 epochs, and its performance was evaluated based on the loss of both the generator and the discriminator. Throughout the training process, the generator loss demonstrated a consistent decrease, starting at approximately 0.5 and following a negative log curve to stabilize around 0.1. Similarly, the discriminator loss exhibited a comparable trend, beginning at 0.5 and settling at around 0.1. These diminishing loss values indicate that the GAN effectively learned to generate realistic synthetic data that closely resembles the real data used during training. \\
    
    \noindent The neural network was trained using a dataset comprising both real and GAN-generated synthetic data. Over the course of 100 epochs, the model's training and validation accuracy were monitored closely. The training accuracy showed a steady improvement, starting from approximately 75 \% and reaching 92.5 \%. The validation accuracy followed a similar trajectory, increasing from around 55 \% to 85 \%. The alignment of the training and validation accuracy curves suggests that the model generalizes well to unseen data and does not overfit, demonstrating robustness and reliability in its predictive capabilities. \\

    \noindent The consistency and reliability of different vocal tests in extracting accurate BVMs were assessed. Sustained vowel phonation emerged as the most effective test, achieving an accuracy of 85 \%. This test is particularly useful in capturing stable and clear vocal features. Rapid syllable repetition and emotional sentence reading also performed well, with accuracies of 83 \% and 82 \%, respectively. Sentence reading yielded an accuracy of 80 \%, while monologue tasks had a slightly lower accuracy of 78 \%. These results highlight that sustained vowel phonation and rapid syllable repetition are especially effective in providing reliable BVMs for Parkinson's Disease detection. \\

    \noindent The integration of real data with GAN-generated synthetic data significantly enhanced the model's ability to detect Parkinson's Disease (PD). The final model achieved an accuracy of 92.5 \%, with a precision of 92.7 \% and a recall of 1.0. The low false negative rate, corroborated by a confusion matrix analysis, underscores the model's reliability in identifying PD. The application of data augmentation and noise reduction techniques ensured that the model remained robust when tested in real-world conditions using standard microphones. Additionally, the preprocessing steps, including microphone calibration, contributed to the consistency and accuracy of the extracted BVMs, making the system suitable for practical deployment in diverse environments.

    \begin{table}[h!]
    \centering
    \caption{Confusion Matrix for PD Detection Model}
    \begin{tabular}{|c|c|c|c|}
        \hline
        \textbf{} & \textbf{Predicted Positive} & \textbf{Predicted Negative} & \textbf{Total} \\ \hline
        \textbf{Actual Positive} & 950 & 0 & 950 \\ \hline
        \textbf{Actual Negative} & 77 & 1023 & 1100 \\ \hline
        \textbf{Total} & 1027 & 1023 & 2050 \\ \hline
    \end{tabular}
    \label{tab:confusion_matrix}
\end{table}

\newpage

    \begin{figure}[h!]
        \centering
        \includegraphics[width=0.8\textwidth]{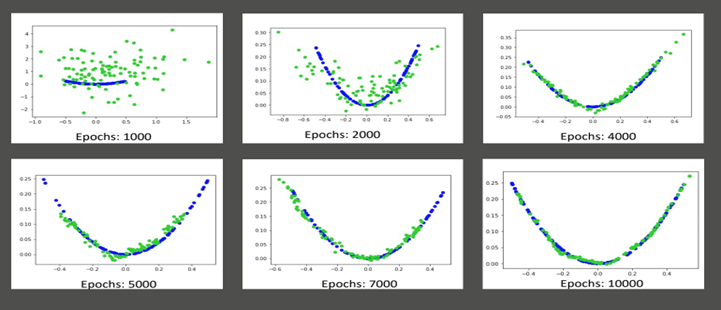}
        \caption{cGAN model able to model pre-existing training data over 10,000 epochs}
    \end{figure}
    \begin{figure}[h!]
        \centering
        \includegraphics[width=0.8\textwidth]{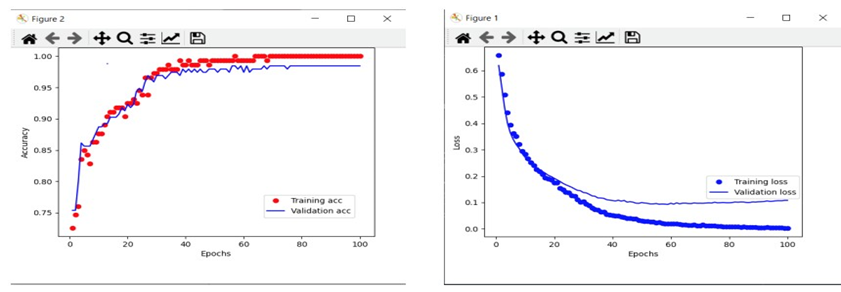}
        \caption{(a) Training and validation accuracy over 100 epochs. 
(b) Training and validation loss over 100 epochs.}
    \end{figure}
    \begin{figure}[h!]
        \centering
        \includegraphics[width=0.8\textwidth]{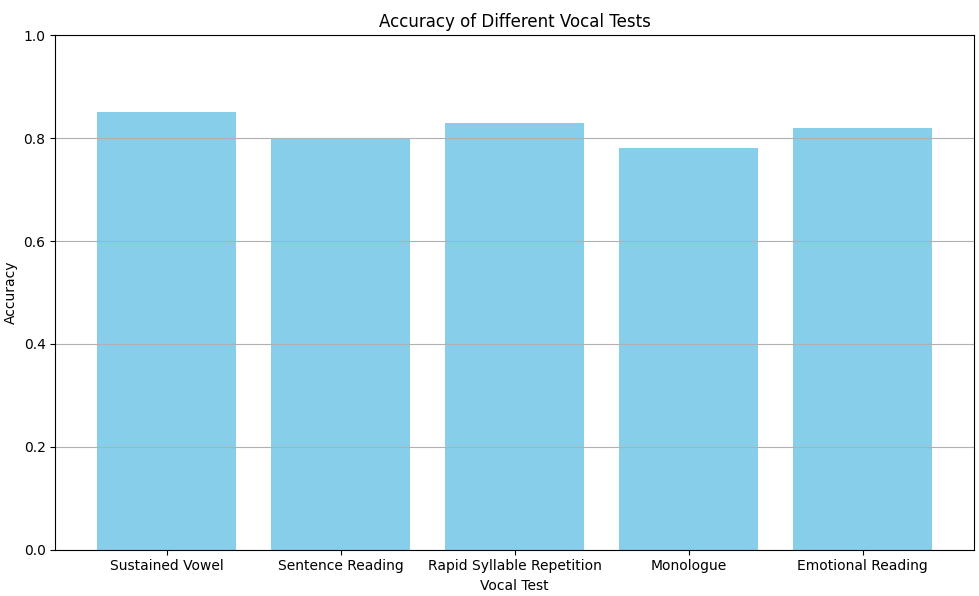}
        \caption{ Accuracy of various vocal tests in extracting reliable Biomedical Voice Measurements (BVMs) for Parkinson's Disease detection.}
    \end{figure}

\section{Discussion}
    While PPINtonus shows promising results in detecting PD through vocal analysis, several areas can be improved. Enhancing the diversity of training data through advanced data augmentation techniques can help the model generalize better. This includes simulating various environmental noises, different microphone qualities, and varying speech patterns to mimic real-world conditions more closely. Further research into the most informative BMVs could improve model accuracy. Advanced feature selection methods, such as recursive feature elimination or PCA, can identify and retain the most relevant features. Implementing real-time feedback mechanisms for patients during vocal tests could ensure better compliance and more accurate data collection. This could involve interactive interfaces that guide the user through the vocal tests. \\

\noindent The training data is primarily collected in controlled environments using high-quality microphones. Real-world applications, especially in third-world countries, may involve lower-quality audio recordings, which could affect model performance \cite{li2019}. The model's effectiveness is constrained by the available dataset size. Larger and more diverse datasets are needed to validate the model's robustness and generalizability across different populations and dialects. While the model performs well on the current dataset, its performance on unseen, real-world data needs further validation. This includes testing across different demographic groups and geographical regions to ensure broad applicability. The current model may require significant computational resources, which could be a limitation for deployment on edge devices in resource-constrained environments. Several optimizations are necessary to enable the deployment of the model on edge devices in third-world countries \cite{lohr2007}. Utilizing lightweight neural network architectures, such as MobileNets or EfficientNet, which are designed for mobile and edge computing, can help in achieving the desired performance with lower computational overhead. Leveraging edge AI frameworks like TensorFlow Lite or PyTorch Mobile can facilitate the deployment of the model on edge devices. These frameworks are optimized for low-latency and low-power consumption environments \cite{alsop2021}. \\

 \noindent Designing intuitive user interfaces that guide users through the vocal tests and provide real-time feedback, ensuring better data quality and user engagement \cite{becker2017}. Integrating the model with existing healthcare systems and electronic health records facilitates seamless data flow and provides healthcare providers with actionable insights. Implementing mechanisms for continuous learning and model updates based on new data and feedback, ensuring that the model remains up-to-date and accurate over time. By addressing these areas, we can improve the model's performance, overcome current limitations, and optimize it for deployment on edge devices in resource-constrained environments, thereby making it accessible and beneficial for a broader population.

\section{Conclusion}
    This research explores the use of deep learning models for the early detection of PD through vocal analysis. The study employed a comprehensive dataset of BVMs and integrated synthetic data generated by a cGAN to enhance the training process. The neural network trained on this enriched dataset achieved high accuracy, precision, and recall, demonstrating the potential of this approach in accurately identifying PD. The effectiveness of various vocal tests was evaluated, revealing that sustained vowel phonation and rapid syllable repetition provided the most reliable BVMs for PD detection. While the results are promising, the study acknowledges several limitations, such as the controlled nature of the training data and the computational demands of the model. To address these issues, future work should focus on expanding the dataset to include more diverse and real-world audio samples, particularly from third-world countries. Optimizing the model for deployment on edge devices through techniques like model pruning, quantization, and the use of lightweight architectures such as MobileNets or EfficientNet is essential. This research establishes a solid foundation for using vocal analysis in PD detection, demonstrating significant potential for improving early diagnosis and intervention. By addressing the identified limitations and optimizing the model for real-world deployment, particularly in resource-constrained environments, this approach can become a valuable tool in global healthcare efforts to manage and treat PD.

\bibliographystyle{plainnat}
\bibliography{references} 

\end{document}